\documentclass[12pt]{article}

\usepackage{latexsym}
\usepackage{graphicx}
\usepackage{verbatim}
\usepackage{amssymb,amsfonts,amscd,amsbsy}

\catcode`\@=11
\def\marginnote#1{}
\newcount\hour
\newcount\minute
\newtoks\amorpm
\hour=\time\divide\hour by60
\minute=\time{\multiply\hour by60 \global\advance\minute by-\hour}
\edef\standardtime{{\ifnum\hour<12 \global\amorpm={am}%
        \else\global\amorpm={pm}\advance\hour by-12 \fi
        \ifnum\hour=0 \hour=12 \fi
        \number\hour:\ifnum\minute<10 0\fi\number\minute\the\amorpm}}
\edef\militarytime{\number\hour:\ifnum\minute<10 0\fi\number\minute}
\def\draftlabel#1{{\@bsphack\if@filesw {\let\thepage\relax
   \xdef\@gtempa{\write\@auxout{\string
      \newlabel{#1}{{\@currentlabel}{\thepage}}}}}\@gtempa
   \if@nobreak \ifvmode\nobreak\fi\fi\fi\@esphack}
        \gdef\@eqnlabel{#1}}
\def\@eqnlabel{}
\def\@vacuum{}
\def\draftmarginnote#1{\marginpar{\raggedright\scriptsize\tt#1}}
\def\draft{\oddsidemargin -.5truein
        \def\@oddfoot{\sl preliminary draft \hfil
        \rm\thepage\hfil\sl\today\quad\militarytime}
        \let\@evenfoot\@oddfoot \overfullrule 3pt
        \let\label=\draftlabel
        \let\marginnote=\draftmarginnote
   \def\@eqnnum{(\theequation)\rlap{\kern\marginparsep\tt\@eqnlabel}%
\global\let\@eqnlabel\@vacuum}  }

\def\preprint{\twocolumn\sloppy\flushbottom\parindent 1em
        \leftmargini 2em\leftmarginv .5em\leftmarginvi .5em
        \oddsidemargin -.5in    \evensidemargin -.5in
        \columnsep 15mm \footheight 0pt
        \textwidth 250mmin      \topmargin  -.4in
        \headheight 12pt \topskip .4in
        \textheight 175mm
        \footskip 0pt
        \def\@oddhead{\thepage\hfil\addtocounter{page}{1}\thepage}
        \let\@evenhead\@oddhead \def\@oddfoot{} \def\@evenfoot{} }

\def\titlepage{\@restonecolfalse\if@twocolumn\@restonecoltrue\onecolumn
     \else \newpage \fi \thispagestyle{empty}\c@page\z@ 
        \def\thefootnote{\fnsymbol{footnote}} }

\def\endtitlepage{\if@restonecol\twocolumn \else  \fi
        \def\thefootnote{\arabic{footnote}}
        \setcounter{footnote}{0}}  

\catcode`@=12
\relax

\def\bea{\begin{array}}
\def\bem{\begin{displaymath}}
\def\beq{\begin{equation}}

\def\eea{\end{array}}
\def\eem{\end{displaymath}}
\def\eeq{\end{equation}}

\def\Im{\mathop{\rm Im}}

\def\ov{\overline}

\def\Re{\mathop{\rm Re}}

\def\s2w{\sin^2 \theta_W}
\def\Tr{\mathop{\rm Tr}}

\relax
\def\dalpha{{\dot\alpha}}

\def\crbig{\\\noalign{\vspace {3mm}}}
\def\bigint{{\displaystyle\int}}

\def\L{{\cal L}}

\def\Dint{\bigint d^2\theta d^2\ov\theta\,}
\def\Fint{\bigint d^2\theta\,}
\def\Fbarint{\bigint d^2\ov\theta\,}

\newcommand{\skipthispart}[1]{}

\relax

\begin{document}
\topmargin-2.4cm
%
%
%
%
\begin{titlepage}
\begin{flushright}
September 2, 2016
\end{flushright}
\vspace{1.3cm}

\begin{center}{\Large\bf
Currents in supersymmetric field theories   }
\vspace{1.3cm}

{\large\bf Jean-Pierre Derendinger\,\footnote{\,derendinger@itp.unibe.ch}  
}
\vskip 1.1cm
Albert Einstein Center for Fundamental Physics \\
Institute for Theoretical Physics, University of Bern \\
Sidlerstrasse 5, CH--3012 Bern, Switzerland \\
\end{center}
\vspace{1.3cm}

\begin{center}
{\large\bf Abstract}
\end{center}
\begin{quote}
A general formalism to construct and improve supercurrents and source or anomaly superfields in two-derivative
${\cal N}=1$ supersymmetric theories is presented. It includes arbitrary gauge and chiral superfields
and a linear superfield coupled to gauge fields. These families of supercurrent structures are characterized by their
energy-momentum tensors and $R$ currents and they display a specific relation to the dilatation current of the
theory. The linear superfield is introduced in order to describe the gauge coupling as a background (or
propagating) field.  Supersymmetry does not constrain the dependence on this gauge coupling field 
of gauge kinetic terms and holomorphicity restrictions are absent. Applying these results to an effective (Wilson) 
description of super-Yang-Mills
theory, matching or cancellation of anomalies leads to an algebraic derivation of the all-order NSVZ $\beta$ function.
\end{quote}

\vspace{1.2cm}

\begin{center}
{\it Invited contribution to Planck 2015, From the Planck Scale to the Electroweak Scale, 25--29 May 2015, 
Ioannina, Greece}
\end{center}

\end{titlepage}
\setcounter{footnote}{0}
\setcounter{page}{0}
\setlength{\baselineskip}{.6cm}
\setlength{\parskip}{.2cm}
\newpage
\renewcommand{\theequation}{\thesection.\arabic{equation}}

%
\section{Introduction: relativistic field theories}

Currents are in direct relation with the algebras of transformations acting on fields and on the action.
For on-shell fields, currents verify a conservation equation giving their divergence in terms
of the variation of the Lagrangian ${\cal L}$ under the associated field and coordinate variations. 
For exact symmetries, currents are then conserved, but significant information also arises 
from variations which are not symmetries, if the lagrangian variation is understood. 
Familiar examples are chiral and scale transformations which could be classical symmetries of massless 
field theories but are violated by calculable quantum anomalies.

Relativistic field theories have Poincar\'e symmetry. Fields transform linearly in a representation characterized
by generators 
\beq
\label{Poin1}
P^\mu = -i \, \partial^\mu \quad\makebox{(translations)},
\qquad\quad
M^{\mu\nu} = \Sigma^{\mu\nu} + i \, x^\mu\partial^\nu - i \, x^\nu\partial^\mu
\quad\makebox{(Lorentz)},
\eeq
verifying the Poincar\'e algebra
\beq
\label{Poin2}
\begin{array}{rcl} 
[M^{\mu\nu},M^{\rho\sigma}]&=& -i\left( \eta^{\mu\rho}M^{\nu\sigma}+ 
\eta^{\nu\sigma}M^{\mu\rho}-\eta^{\mu\sigma}M^{\nu\rho} 
-\eta^{\nu\rho}M^{\mu\sigma} \right) ,  \crbig
[ M^{\mu\nu} , P^\rho ] &=& -i\left( \eta^{\mu\rho}P^\nu -\eta^{\nu\rho}P^\mu\right), 
\qquad\qquad
[P^\mu,P^\nu] \,\,=\,\,  0.
\end{array}
\eeq
Hence, the Poincar\'e properties of fields are encoded in the choice of Lorentz generators 
$\Sigma_{\mu\nu}$. There are ten conserved currents: four translation currents assembled in the in 
general non-symmetric energy-momentum tensor $t_{\mu\nu}$, $\partial^\mu t_{\mu\nu}=0$ and six 
Lorentz currents $j_{\mu,\nu\rho} = - j_{\mu,\rho\nu}$, $\partial^\mu j_{\mu,\nu\rho}=0$. 
But Lorentz symmetry can be used to eliminate the six antisymmetric components of the
energy-momentum tensor, to obtain the symmetric Belinfante tensor $T_{\mu\nu}$. Lorentz currents
read then
\beq
\label{Belinf1}
j_{\rho,\mu\nu}  = - x_\mu T_{\rho\nu} + x_\nu T_{\rho\mu}
\eeq
and the generators $\Sigma_{\mu\nu}$ only appear in the construction of $T_{\mu\nu}$. If the theory
is coupled with diffeomorphism invariance to a background metric $g_{\mu\nu}$ or to a vierbein $e_\mu^a$, the Belinfante tensor
is also obtained as 
\beq
\label{Belinf2}
T_{\mu\nu} = {2\over e} {\partial{\cal L}\over\partial g^{\mu\nu}}
= {1\over2e}\left[ {\partial{\cal L}\over\partial e^\mu_a} \, e_{\nu a}
+  {\partial{\cal L}\over\partial e^\nu_a} \, e_{\mu a} \right] .
\eeq

There are two relevant extensions of Poincar\'e space-time symmetry: firstly scale transformations (or dilatations)
of fields and coordinates with algebra
\beq
\label{scale1}
[ M_{\mu\nu} , D ] = 0,
\qquad\qquad
[ D , P_\mu  ] = i P^\mu .
\eeq
The second relation indicates that we count scale dimensions in energy units ($P_\mu$ has scale dimension 
$+1$). Variations are 
\beq
\label{scale2}
\delta x^\mu = -\lambda x^\mu, \qquad \qquad \delta\phi = i \lambda D\phi, \qquad
D = -i \, x^\mu\partial_\mu - i \, {\cal D} .
\eeq
They are defined by assigning scale dimensions in matrix ${\cal D}$ (or its eigenvalues $w$) 
to fields or operators. The scale or dilatation current depends on these scale dimensions:
\beq
\label{scale3}
j_\mu^D = {\cal V}_\mu + x^\nu T_{\mu\nu},
\eeq
in terms of the Belinfante energy-momentum tensor, with {\it virial current}
\beq
\label{scale4}
{\cal V}_\mu =  {\partial{\cal L}\over\partial\partial_\rho\Phi}(\eta_{\mu\rho}{\cal D}\Phi+ i\Sigma_{\rho\mu}\Phi ).
\eeq
Once Lorentz generators $\Sigma_{\mu\nu}$ and scale dimensions ${\cal D}$ have been assigned to fields,
conformal boost variations with generators
\beq
\label{conf1}
K^{\mu} = -i (2x^\mu x^\nu -  \eta^{\mu\nu}x^2)\partial_\nu - 2i x^\mu{\cal D}
- 2 \Sigma^{\mu\nu} x_\nu
\eeq
follow. The conformal algebra $SO(2,4)\sim SU(2,2)$ is completed by
\beq
\label{conf2}
\begin{array}{rcl}
[ M^{\mu\nu} , K^\rho ] &=& -i\left( \eta^{\mu\rho}K^\nu -\eta^{\nu\rho}K^\mu\right),
\qquad\qquad
[ K^\mu , K^\nu ] \,\,=\,\, 0 ,
\crbig
[ P^\mu , K^\nu ]Ê&=& -2i\, ( \eta^{\mu\nu} D + M^{\mu\nu} ) ,
\hspace{2.07cm}
[ D , K^\mu ] \,\,=\,\, - i K^\mu .
\end{array}
\eeq
Since the four currents of conformal boosts (or special conformal transformations) can be expressed as
\beq
\label{conf3}
K_\rho^\mu = 2 x^\mu j_\rho^D - x^2\, {T_\rho}^\mu,
\eeq
the conservation equations for the dilatation and conformal currents are
\beq
\label{conf4} 
\partial^\mu j_\mu^D = \partial^\mu{\cal V}_\mu + {T^\mu}_\mu \, , \qquad\qquad\qquad
\partial^\rho K_\rho^\mu = 2 x^\mu \partial^\rho j_\rho^D + 2 \, {\cal V}^\mu.
\eeq
In the second equation, ${\cal V}_\mu$ is the virial current (\ref{scale4}) associated with the Belinfante energy-momentum tensor $T_{\mu\nu}$.

Invariance under special conformal transformations generated by (\ref{conf1}) requires scale invariance.
This follows already from the third commutator (\ref{conf2}). Scale invariance implies full conformal symmetry if
the virial current is a derivative, ${\cal V}_\mu = \partial^\nu\sigma_{\mu\nu}$ \cite{CCJ, CJ, C, P}. 
In this case, one can replace the currents $K^\mu_\rho$ by 
\beq
\label{conf5}
\widehat K^\mu_\rho = K^\mu_\rho - 2\,{\sigma^\mu}_\rho \qquad \makebox{with} \qquad 
\partial^\rho\widehat K_\rho^\mu = 2 x^\mu \, \partial^\rho j_\rho^D.
\eeq
Or one can improve the Belinfante energy-momentum tensor to the Callan-Coleman-Jackiw (CCJ) tensor
$\Theta_{\mu\nu}$, to eliminate the virial current and obtain
\beq
\label{conf6}
j_\mu^D = x^\nu\,\Theta_{\mu\nu}, \qquad\qquad
\partial^\rho K^\mu_\rho = 2x^\mu \, {\Theta^\rho}_\rho.
\eeq
Tracelessness of the CCJ tensor implies then conformal symmetry. In general, the first equation 
(\ref{conf6}) defines the CCJ energy-momentum tensor. It exists if ${\cal V}_\mu = 
\partial^\nu\sigma_{\mu\nu}$. The violation or conservation of scale symmetry is measured by the trace
of the energy-momentum tensor in this case only.

\section{Supersymmetric field theories}
\setcounter{equation}{0}

Supersymmetry\footnote{We consider here ${\cal N}=1$ supersymmetry only.} extends the Poincar\' e 
algebra with spin $1/2$ generators $Q_\alpha$ and
$\ov Q_\dalpha = Q_\alpha^\dagger$: 
\beq
\label{susy1}
\begin{array}{rclrcl}
[ M_{\mu\nu} , Q_\alpha] &=& \displaystyle -{i\over4} \, ( [\sigma_\mu,\ov\sigma_\nu]Q )_\alpha , \qquad\qquad&
[P_\mu, Q_\alpha] &=& 0,
\crbig
\{ Q_\alpha , \ov Q_\dalpha \} &=& 2 \, {\sigma^\mu}_{\alpha\dalpha} P_\mu ,
\qquad&
\{ Q_\alpha , Q_\beta \} &=& 0.
\end{array}
\eeq
The corresponding conserved {\it supercurrent} $S_{\mu\alpha}$, $\partial^\mu S_{\mu\alpha}=0$ has
$16_F-4_F=12_F$ operator components. Since currents are themselves local fields, they transform under
Poincar\' e ($T_{\mu\nu}$ is a two-tensor, $S_{\mu\alpha}$ is a spinor-vector) and also under
supersymmetry. It is natural to expect that currents assemble in supermultiplets, requiring however
 an equal number of bosonic and fermionic components. This cannot be achieved with the
$6_B$ components of the symmetric energy-momentum tensor characterizing Poincar\'e symmetry.

Supermultiplets of ${\cal N}=1$ Poincar\'e supersymmetry are also representations of the superconformal 
${\cal N}=1$ superalgebra $SU(2,2|1)$, with bosonic sector $SU(2,2)\times U(1)_R \sim SO(2,4) 
\times  U(1)_R$. One simply needs to assign a scale dimension $w$ (as in the conformal case)
and a $U(1)_R$ charge $q$ to each component field in the theory to fully define the superconformal variations. 
Normalizing $U(1)_R$ with\footnote{Another convention exists in the literature, with $3/2$ replaced by $1$.}
\beq
[ R , Q_\alpha ] = -{3\over2} i Q_\alpha, \qquad\qquad
[ R , \ov Q_\dalpha ] = {3\over2} i \ov Q_\dalpha, 
\eeq
there are three simple rules: a chiral superfield has $w=q$, a real linear superfield\footnote{See below.} 
has $w=2$, $q=0$ and
of course a real superfield has $q=0$. It follows that the chiral superfield of gauge field strengths ${\cal W}_\alpha$ 
has $R$--charge $q=3/2$. Notice that $U(1)_R$ charge assignments can always be applied in Poincar\'e 
supersymmetry since chiral  multiplet scalars 
live on a K\"ahler manifold. But $U(1)_R$ is not a symmetry in general, and it is not uniquely defined. 

The structure of currents is as follows. Firstly, conformal invariance can be summarized in
the existence of a conserved, symmetric, traceless (CCJ) energy-momentum tensor $\Theta_{\mu\nu}$ with $5_B$ 
fields. Secondly, $U(1)_R$ symmetry implies the existence of a conserved current ${\cal J}_\mu$ 
with $3_B$ fields. Thirdly, $SU(2,2|1)$ has eight supersymmetry generators. 
The supplementary (with respect to the Poincar\'e case)
special supersymmetry allows to remove the ``$\gamma$--trace" of the supercurrent $S_{\mu\alpha}$:
$(\ov\sigma^\mu S_\mu)^\dalpha=0$, and $8_F$ fields remain in the supercurrent.\footnote{
This is somewhat similar to Lorentz symmetry, used to symmetrize the energy-momentum tensor.}
Hence, the energy-momentum
tensor, the $R$--current and the supercurrent include a total of $8_B+8_F$ operators. 
\begin{table}[ht]
\begin{center}
\framebox{
\begin{tabular}{ llll }
Poincar\'e: & $6_B+12_F$ \qquad & \quad $\partial^\mu T_{\mu\nu}= \partial^\mu S_{\mu\alpha} = 0$ 
& \quad $T_{\mu\nu}=T_{\nu\mu}$
\crbig
Conformal: & $ 8_B+8_F$ \qquad & \quad $\partial^\mu \Theta_{\mu\nu} = \partial^\mu S_{\mu\alpha} = \partial^\mu{\cal J}_\mu =0$
& \quad $\Theta_{\mu\nu}=\Theta_{\nu\mu}$ 
\crbig
&&&  \quad
${\Theta^\mu}_\mu= (\ov\sigma^\mu S_\mu)^\dalpha = 0$
\end{tabular}
}
\caption{Current structure of ${\cal N}=1$ theories}
\end{center}
\end{table}

In 1975, Ferrara and Zumino \cite{FZ} showed that in conformal Wess-Zumino and in super-Yang-Mills 
(classical) theories, the currents $\Theta_{\mu\nu}$ and $S_{\mu\alpha}$
belong to a supermultiplet with an appropriate $R$--symmetry current 
${\cal J}_\mu$. The supermultiplet is cast in the real superfield
\beq
\label{4scdef1}
J_\mu = (\ov\sigma_\mu)^{\dalpha\alpha}\, J_{\alpha\dalpha} \, , \qquad\qquad\qquad
J_{\alpha\dalpha}= {1\over2}(\sigma^\mu)_{\alpha\dalpha} \, J_\mu 
\eeq 
submitted for on-shell fields to the supercurrent superfield equation 
\beq
\ov D^\dalpha J_{\alpha\dalpha} = 0
\eeq
which includes all conservation laws and trace conditions.
They also showed that breaking conformal symmetry with superpotential terms in the Wess-Zumino model 
introduces a specific source term in the superfield equation, $\ov D^\dalpha J_{\alpha\dalpha} = \Delta_\alpha$,
generates values for ${\Theta^\mu}_\mu$, $\partial^\mu{\cal J}_\mu$, $(\ov\sigma^\mu S_\mu)^\dalpha$
and also adds $4_B+4_F$ fields in the supercurrent superfield structure, to obtain $12_B+12_F$ 
fields.\footnote{As in the off-shell supermultiplet of minimal ${\cal N}=1$ supergravity.}

In the superconformal case, the assignments of $R$--charges and scale dimensions are constrained by superconformal symmetry. In contrast, in Poincar\'e supersymmetry, these numbers are mostly arbitrary if 
no choice leads to scale or 
$R$ invariance. The $R$ and dilatation currents depend on these numbers and, since their supersymmetry partners
$T_{\mu\nu}$ and $S_{\mu\alpha}$ do not depend on $q$ or $w$, the corresponding supermultiplet of 
currents will include a $U(1)_R$ current with specific $R$--charges.

\section{Supercurrent structures}
\setcounter{equation}{0}

This section discusses the supercurrent superfields and equations relevant for arbitrary two-derivative
${\cal N}=1$ lagrangian field theories, following refs.~\cite{ADH, AADH} and also borrowing several results 
from ref.~\cite{KS}.

\subsection{The supercurrent superfield equation}

As originally shown by Ferrara and Zumino \cite{FZ}, the conserved supercurrent 
$\partial^\mu S_{\mu\alpha}=0$ can be embedded in a real Lorentz vector superfield
$J_{\alpha\dalpha}$ submitted to a differential superfield supercurrent equation when fields solve field equations. 
At this point, $J_{\alpha\dalpha}$ includes $32_B+32_F$ components, or $8_B+8_F$ currents.
One needs to impose a superfield differential equation to impose current conservation and reduce the 
number of components.
The supercurrent equation is actually of the form\footnote{The conjugate of
$\ov D^\dalpha J_{\alpha\dalpha}$ is $-D^\alpha J_{\alpha\dalpha}$.}
\beq
\label{4sc2}
\ov D^\dalpha\, J_{\alpha\dalpha} = \Delta_\alpha \, , \qquad\qquad\qquad
D^\alpha\, J_{\alpha\dalpha} = - \ov\Delta_\dalpha \, ,
\eeq
which implies\index{Supercurrent equation}
\beq
\label{4sc3}
\ov{DD}\,\Delta_\alpha = 0 \, , \qquad\qquad 
D^\alpha\Delta_\alpha + \ov D_\dalpha\ov\Delta^\dalpha = -2i \, \partial^\mu J_\mu \, .
\eeq
The complex linear spinor superfield $\Delta_\alpha$ is the source of the 
non-conser\-va\-tion of (some of) the currents in $J_\mu$. But $\Delta_\alpha$ is not an arbitrary 
linear superfield: it should be such that $J_\mu$ submitted to the supercurrent equation (\ref{4sc2}) 
includes the conserved energy-momentum tensor and supercurrent required 
by super-Poincar\'e invariance of the theory. 

For all supersymmetric field theories considered here, the source or anomaly superfield $\Delta_\alpha$ 
verifying this condition is of the form\footnote{These sources in the supercurrent equation are not the 
most general allowing conserved energy-momentum tensor and supercurrent. See also 
refs.~\cite{CPS, MSW, Kuz}. For a long time, the literature propagated an unfortunate claim \cite{CPS} that 
the coexistence of $\chi_\alpha$ and $X$ is forbidden. 
Refs.~\cite{MSW, Osb} and more explicitly \cite{KS} have removed the ban.}
\beq
\label{4sc4}
\begin{array}{lll}
\Delta_\alpha = D_\alpha X + \chi_\alpha \, , \qquad\quad
&\ov\Delta_\dalpha = -\ov D_\dalpha\ov X + \ov\chi_\dalpha \, , 
\quad\qquad &\ov D_\dalpha X = 0 \, , 
\crbig
\chi_\alpha = -{1\over4}\ov{DD}D_\alpha\, U, \qquad\quad
&\ov\chi_\dalpha = {1\over4}DD\ov D_\dalpha\, U, \qquad\quad
&U=U^\dagger,
\end{array}
\eeq
which is certainly linear, $\ov{DD}\Delta_\alpha=0$. Then,
\beq
\label{4sc5}
\{ D^\alpha , \ov D^\dalpha \} J_{\alpha\dalpha} = D^\alpha\Delta_\alpha 
+ \ov D_\dalpha\ov\Delta^\dalpha = DD\, X - \ov{DD}\,\ov X  ,
\eeq
since $\chi_\alpha$, which has the same structure as the Maxwell field strength superfield 
${\cal W}_\alpha$, verifies Bianchi identity $D^\alpha\chi_\alpha = - \ov D_\dalpha\ov\chi^\dalpha$.
Hence, $\chi_\alpha$ does not contribute to $\partial^\mu J_\mu$. 

In total, superfields $J_{\alpha\dalpha}$, $X$ and $\chi_\alpha$ include $40_B+40_F$ real (or
hermitian) components.
Since the supercurrent superfield equation is complex linear,
it imposes $2\times(12_B+12_F)$ conditions on the $40_B+40_F$ components to leave a 
solution expressed in terms of $16_B+16_F$ fields. 

For a given supersymmetric lagrangian, one can derive superfields $J_{\alpha\dalpha}$, $X$ and
$\chi_\alpha$ (or $U$) verifying the supercurrent equation (\ref{4sc2}). These superfields are not unique, there
exists supersymmetric improvement transformations acting on the conserved currents 
$T_{\mu\nu}$ and $S_{\mu\alpha}$ and transforming all other components of the superfields.
We use the terminology {\it supercurrent structure}
\index{Supercurrent structure} for each triplet of superfields $J_{\alpha\dalpha}$, $X$ and $\chi_\alpha$
submitted to the supercurrent equation $\ov D^\dalpha J_{\alpha\dalpha}=D_\alpha X + \chi_\alpha$.

The supercurrent equation (\ref{4sc2}) holds for
solutions of the field equations only. This is similar to Noether currents associated with continuous 
symmetries: their expression follows from the lagrangian (they can be expressed in terms of off-shell fields)
but their conservation holds for solutions of the field equations. 

\subsection{Component expansion}

To display the content of the supercurrent equation 
\beq
\label{4sc6}
\ov D^\dalpha\,J_{\alpha\dalpha} = D_\alpha X + \chi_\alpha, \qquad
\qquad \ov D_\dalpha X=0,\qquad
\chi_\alpha = -{1\over4}\ov{DD}D_\alpha\, U,
\eeq
its component form is needed.
We use the following expansion of the chiral superfields $X$ and $\chi_\alpha$:
\beq
\label{4sc7}
\begin{array}{rcl}
X(y,\theta) &=& x + \sqrt2\,\theta\psi_X - \theta\theta\, f_X \, ,
\crbig
\chi_\alpha(y,\theta) &=& -i \lambda_\alpha + \theta_\alpha \, D 
+ {i\over2}(\theta\sigma^\mu\ov\sigma^\nu)_\alpha F_{\mu\nu}
- \theta\theta\, (\sigma^\mu\partial_\mu\ov\lambda)_\alpha\end{array}
\eeq
in chiral coordinates or
\beq
\begin{array}{rcl}
 X &=& x + \sqrt2\,\theta\psi_X - \theta\theta\, f_X-i\theta\sigma^\mu\bar\theta\partial_\mu x-\frac{i}{\sqrt{2}}\theta\theta\bar\theta\bar\sigma^\mu\partial_\mu\psi_X-\frac{1}{4}\theta\theta\ov{\theta\theta}\Box x ,
 \crbig
 \chi_\alpha &=& -i \lambda_\alpha + \theta_\alpha \, D + \frac{i}{2}(\theta\sigma^\mu\ov\sigma^\nu)_\alpha F_{\mu\nu} - \theta\sigma^\mu\bar\theta\partial_\mu\lambda_\alpha-\theta\theta(\sigma^\mu\partial_\mu\ov\lambda)_\alpha\nonumber
 \crbig
 &&- \frac{1}{2}\theta\theta(\sigma^\mu\bar\theta)_\alpha(\partial^\nu F_{\nu\mu}
- i\partial_\mu D)+\frac{i}{4}\theta\theta\ov{\theta\theta}\Box\lambda_\alpha
\end{array}
\eeq
in ordinary coordinates $(x,\theta,\ov\theta)$.
For the real superfield $U$, the last eq.~(\ref{4sc6}) implies
\beq
\label{4sc7b}
U = \theta\sigma^\mu\ov\theta\, U_\mu + i \,\theta\theta\ov{\theta\lambda} 
+ i \, \ov{\theta\theta}\theta\lambda
+ {1\over2}\theta\theta\ov{\theta\theta}\, D + \ldots
\eeq
where the dots denote components of $U$ absent from $\chi_\alpha$ and
$F_{\mu\nu}=\partial_\mu U_\nu - \partial_\nu U_\mu$. 

With these component expansions, the resulting supercurrent superfield is\footnote{
In the expansion, the normalizations of $j_\mu$ and $T_{\mu\nu}$ have been selected to correspond to
well-defined currents, see below. This has not been done for the supercurrent $S_{\mu\alpha}$ which is not 
explicitly used here. This expansion, originally given in ref.~\cite{KS} with slightly different conventions, 
is not unique, see conditions (\ref{4sc10}). }
\beq
\label{4sc8}
\begin{array}{rcl}
J_\mu (x,\theta,\ov\theta) &=& {8\over3} \, j_\mu(x) 
+ \theta (S_\mu + 2\sqrt2\, \sigma_\mu\ov\psi_X )
+ \ov\theta (\ov S_\mu - 2\sqrt2\, \ov\sigma_\mu\psi_X )
\crbig
&& - 2i \,\theta\theta\,\partial_\mu\ov x + 2i \,\ov{\theta\theta}\, \partial_\mu x
\crbig
&& \displaystyle
+ \theta\sigma^\nu\ov\theta\Bigl( 8\,T_{\mu\nu} - 4\,\eta_{\mu\nu}\Re f_X
- {1\over2}\epsilon_{\mu\nu\rho\sigma}( {8\over3} \,  \partial^\rho j^\sigma - F^{\rho\sigma} )\Bigr)
\crbig
&& \displaystyle - {i\over2} \theta\theta\ov\theta ( \partial_\nu S_\mu \sigma^\nu 
+ 2 \sqrt2\, \ov\sigma_\mu \sigma^\nu \partial_\nu\ov\psi_X)
\crbig
&& \displaystyle + {i\over2} \ov{\theta\theta}\theta ( \sigma^\nu \partial_\nu \ov S_\mu
+ 2\sqrt2\, \sigma_\mu \ov\sigma^\nu \partial_\nu\psi_X)
\crbig
&& \displaystyle
- {2\over3}\theta\theta\ov{\theta\theta}\, \Bigl(  2 \, \partial_\mu \partial^\nu j_\nu
- \Box j_\mu \Bigr)
\end{array}
\eeq
with $T_{\mu\nu} = T_{\nu\mu}$. This expression solves the supercurrent equation (\ref{4sc6}) if $T_{\mu\nu}$
and $S_\mu$ are conserved,
\beq
\label{4sc9}
\partial^\mu T_{\mu\nu} = 0, \qquad\qquad
\partial^ \mu S_\mu = 0.
\eeq
Hence, $T_{\mu\nu}$ and $S_\mu$ will be (proportional to) the conserved energy-momentum 
tensor and the supercurrent. In addition, the supercurrent equation (\ref{4sc6}) implies the following  
conditions:
\beq
\label{4sc10}
\begin{array}{c}
4\, {T^\mu}_\mu = D + 6 \Re f_X, \qquad\qquad
\partial^\mu\,j_\mu = -{3\over2} \,\Im f_X, 
\crbig
(\sigma^\mu \ov S_\mu)_\alpha = 6\sqrt2\,\psi_{X\,\alpha} + 2i\,\lambda_\alpha.
\end{array}
\eeq
The first condition indicates that both superfields $X$ and $\chi_\alpha$ are sources for the 
trace of the energy-momentum tensor. Its precise significance depends on the specific 
energy-momentum tensor included in $J_\mu$: since the energy-momentum tensor $T_{\mu\nu}$ is 
defined up to improvements, the relation between the trace ${T^\mu}_\mu$ and scale invariance
or violation in the theory depends on the choice of $T_{\mu\nu}$.
The second condition (\ref{4sc10})
indicates that $X$ only induces the nonconservation of $j_\mu$, which
is related in general to a $R$ transformation acting in the theory. 
The third condition controls the violation of conformal supersymmetry.
Hence, the presence of the source $\chi_\alpha$ breaks the correlation between ${T^\mu}_\mu$
and $\partial^\mu\,j_\mu$.

The scale dimensions of the component fields are:
\beq
\begin{array}{rcll}
3:& \qquad & J_{\alpha\dalpha}\,, \quad X\,; \qquad & j_\mu\,, \quad x\,, \quad U_\mu\,;
\crbig
7/2:& & \chi_\alpha\,; & S_{\mu\alpha}\,, \quad\psi_{X\alpha}\,, \quad \lambda_\alpha\,;
\crbig
4:& & & T_{\mu\nu}\,, \quad f_X \,, \quad D\,, \quad F_{\mu\nu}\,.
\end{array}
\eeq

To see for instance how the conservation of the energy-momentum tensor follows from the supercurrent 
equation, write
$$
\begin{array}{l}
J_\mu = {8\over3} \, j_\mu + 8 \,\theta\sigma^\nu\ov\theta \, t_{\mu\nu} + \theta\theta\ov{\theta\theta}\,d_\mu + \ldots
\crbig
t_{\mu\nu} = T_{\mu\nu} + \eta_{\mu\nu} \, t + \tau_{\mu\nu}, \qquad\qquad
T_{\mu\nu} = T_{\nu\mu}, \qquad \tau_{\mu\nu} = -\tau_{\nu\mu}.
\end{array}
$$
The $\theta_\alpha$ component of the supercurrent equation, after separation of the symmetric and 
antisymmetric parts
$$
\theta\{\sigma^\mu,\ov\sigma^\nu\}_\alpha = 2\,\eta^{\mu\nu}\,\theta_\alpha ,
\qquad\qquad
\theta[\sigma^\mu,\ov\sigma^\nu]_\alpha,
$$
and of real and imaginary parts, provides three equations:\footnote{The real (or imaginary)
antisymmetric part is removed using 
$$
[ \sigma^\mu, \ov\sigma^\nu ] = {i\over2}\epsilon^{\mu\nu\rho\sigma} [ \sigma_\rho , 
\ov\sigma_\sigma ] .
$$ }
$$
\begin{array}{rcl}
{T^\mu}_\mu &=& - 4t + {1\over4}D - {1\over2}\Re f_X  \, , 
\crbig
\partial^\mu j_\mu &=& - {3\over2}\,\Im f_X \, ,
\crbig
\epsilon_{\mu\nu\rho\sigma} \tau^{\rho\sigma} &=&
{1\over3} (\partial_\mu j_\nu - \partial_\nu j_\mu ) - {1\over4}\, F_{\mu\nu} \,.
\end{array}
$$
The (complex) $\theta\theta\ov\theta_\dalpha$ component gives two (real) equations:
$$
\begin{array}{rcl}
d_\mu &=& \partial_\mu\Im f_X - 2\, \epsilon_{\mu\nu\rho\sigma}\partial^\nu\tau^{\rho\sigma} 
- {1\over2}\partial^\nu F_{\mu\nu} \,,
\crbig
\partial^\nu T_{\mu\nu} &=& {1\over2} \, \partial_\mu ( 2 t + {T^\nu}_\nu - {1\over4}\,D - {1\over2}\Re f_X ) \, .
\end{array}
$$
Since then $\partial^\nu T_{\mu\nu} = - \partial_\mu (t + {1\over2} \Re f_X)$, $T_{\mu\nu}$
is conserved if one defines 
$$
t = - {1\over2} \Re f_X .
$$
The five equations provide then the conservation of $T_{\mu\nu}$, the expressions of components 
$\tau_{\mu\nu}$ and $d_\mu$ of $J_\mu$, and the two bosonic constraints (\ref{4sc10}). 

The supercurrent superfield $J_\mu$ 
includes a conserved symmetric energy-momentum tensor $T_{\mu\nu}$ 
($10_B - 4_B = 6_B$), the conserved supercurrent $S_\mu$ ($4\times (4-1)_F = 12_F$) 
and a vector current $j_\mu$ which is not in general conserved ($4_B$). Since conditions (\ref{4sc10})
eliminate $2_B+4_F$, the source superfields $X$ and $\chi_\alpha$ add $6_B+4_F$
fields, for a total of $16_B+16_F$ fields, as earlier mentioned.

Some remarks are in order. 
Firstly, notice that  the components of the anomaly superfields $X$ and $\chi_\alpha$ 
appear in $J_\mu$.  Hence, the symmetric part of the $\theta\sigma^\nu\ov\theta$ component
of $J_\mu$ can only be identified with an energy-momentum tensor of the theory after subtraction of an anomaly contribution generated by $\Re f_X$, or by $D$, or by both,
since we may as well use the first eq.~(\ref{4sc10}) to modify the component expansion (\ref{4sc8}).\footnote{
But it is impossible to eliminate these anomaly contributions from the components $\theta$, $\ov\theta$ and
$\theta\sigma^\nu\ov\theta$  of $J_\mu$ if both $X$ and $\chi_\alpha$ are present. It is apparently the 
{\it assumption} that these components do not depend on the source superfields which led to the claim that $X$ and $\chi_\alpha$ cannot coexist (see eq.~C.5 of ref.~\cite{CPS}).
}

Secondly, even if, for a given theory, one expects to find expressions for $J_{\alpha\dalpha}$, 
$X$ and $\chi_\alpha$ in terms of superfields, {\it i.e.} in terms of off-shell fields,
equations (\ref{4sc8})--(\ref{4sc10}) only hold for on-shell fields. The interpretation of the components of
$J_\mu$ in terms of currents may require the field equations. This is in particular true for the auxiliary 
field contributions.

\subsection{Superfield improvement transformation}

The identity
\beq
\label{4sc11}
2\,\ov D^\dalpha [ D_\alpha , \ov D_\dalpha ] \, {\cal G}
= D_\alpha \, \ov{DD}\,{\cal G} + 3\, \ov{DD} \, D_\alpha \,{\cal G} ,
\eeq
which holds for any superfield ${\cal G}$, is clearly a solution of the supercurrent superfield 
equation (\ref{4sc6}) with $J_{\alpha\dalpha} = 2\,[ D_\alpha , \ov D_\dalpha ] \, {\cal G}$, 
$X= \ov{DD}\,{\cal G}$ and $\chi_\alpha = 3\,\ov{DD} \, D_\alpha \,{\cal G}$ (with ${\cal G}$ real). 
Hence, given superfields $J_{\alpha\dalpha}$, $X$ and $\chi_\alpha$ verifying the supercurrent 
equation, the transformation\index{Superfield improvement transformation}
\beq
\label{4sc12}
\begin{array}{rcl}
J_{\alpha\dalpha} \qquad&\longrightarrow&\qquad \widetilde J_{\alpha\dalpha} = J_{\alpha\dalpha} 
+ 2\,[ D_\alpha , \ov D_\dalpha ] \, {\cal G} , 
\crbig
X \qquad&\longrightarrow&\qquad \widetilde X = X + \ov{DD}\,{\cal G} ,
\crbig
\chi_\alpha \qquad&\longrightarrow&\qquad \widetilde\chi_\alpha = 
\chi_\alpha + 3\, \ov{DD} \, D_\alpha \,{\cal G},
\end{array}
\eeq
is an ambiguity in the realization of the supercurrent superfield. The transformation necessarily involves 
improvement terms for $T_{\mu\nu}$ and $S_{\alpha\mu}$: the transformed supercurrent superfield 
verifies again equation (\ref{4sc6}) and identity (\ref{4sc11}) holds without using any field equation. The 
transformed energy-momentum tensor and supercurrent are then conserved and the modifications 
are improvements. On the other hand, the lowest component $j_\mu$ and the trace ${T^\mu}_\mu$, in 
particular, are non-trivially transformed. 

Hence, each theory admits in principle a (continuous) family of supercurrent structures. Notice that if 
${\cal G}$ is linear ($\ov{DD}{\cal G}= 0$), $\widetilde X=X$. Similarly, if ${\cal G}=\Psi + \ov \Psi$, 
$\ov D_\dalpha\Psi=0$, then $\widetilde\chi_\alpha = \chi_\alpha$. But the use of transformations 
(\ref{4sc12}) may face various obstructions if conditions like gauge invariance or global definition are 
imposed on the supercurrent structure $J_{\alpha\dalpha}$, $X$, 
$\chi_\alpha$.\footnote{Although these superfields are not strictly speaking physical quantities.
These conditions 
have been discussed in ref.~\cite{KS} for some specific theories. See also ref.~\cite{ADH}.}

If the real superfield $\cal G$ of the transformation (\ref{4sc12}) has the expansion
\beq
\begin{array}{rcl}
 \cal G &=& C_g+i\theta\chi_g-i\bar\theta\bar\chi_g+\theta\sigma^\mu\bar\theta v_{g\mu}+\frac{i}{2}\theta\theta(M_g+iN_g)-\frac{i}{2}\ov{\theta\theta}(M_g-iN_g)
\crbig
&& + i\theta\theta\bar\theta(\bar\lambda_g+\frac{i}{2}\partial_\mu\chi_g\sigma^\mu)-i\ov{\theta\theta}\theta(\lambda_g-\frac{i}{2}\sigma^\mu\partial_\mu\bar\chi_g)+\frac{1}{2}\theta\theta\ov{\theta\theta}(D_g-\frac{1}{2}\Box C_g),
\end{array}
\eeq
then the components of the transformed superfields $\widetilde J_\mu$, $\widetilde X$ and $\widetilde\chi_\alpha$ read
\beq
\begin{array}{rclrcl}
\widetilde j_\mu &=& j_\mu-3v_{g\mu} ,
\quad&\quad
\widetilde S_\mu &=& S_\mu+8\sigma_{[\mu}\bar\sigma_{\nu]}\partial^\nu\chi_g,
\crbig
\widetilde\psi_X &=& \psi_X+2\sqrt{2}i\lambda_g+2\sqrt{2}\sigma^\mu\partial_\mu\ov\chi_g,
\quad&\quad
\widetilde x &=& x+2i(M_g-iN_g),
\crbig
\widetilde T_{\mu\nu} &=& T_{\mu\nu}+(\partial_\mu\partial_\nu-\eta_{\mu\nu}\Box) C_g,
\quad&\quad
\widetilde f_X &=& f_X+2D_g-2\Box C_g+2i\partial_\mu v^\mu_g,
\crbig
\widetilde F_{\mu\nu} &=& F_{\mu\nu}-24\partial_{[\mu}v_{g\nu]},
\quad&\quad
\widetilde\lambda &=& \lambda-12\lambda_g,
\crbig
\widetilde D &=& D-12D_g , &&&
\end{array}
\eeq
using the expansions (\ref{4sc7}) and (\ref{4sc8}) of $J_\mu$, $X$ and $\chi_\alpha$.
As expected, the transformations of the energy-momentum tensor $T_{\mu\nu}$ and of the supercurrent 
$S_\mu$ are improvements. 

For a given theory, each supercurrent structure is characterized either by the lowest component $j_\mu$ of
$J_{\alpha\dalpha}$, which is a $U(1)_R$ current, or by the type of energy-momentum tensor it contains.

\subsection{Reductions, coupling to supergavity}

There are three simple reductions of the supercurrent structure with superfields $J_{\alpha\dalpha}$, $X$ and 
$\chi_\alpha$.
Firstly, the {\it Ferrara-Zumino} (FZ) structure \cite{FZ} with $12_B+12_F$ component fields (or operators):
\beq
\makebox{FZ structure:} \qquad\qquad
\chi_\alpha = 0, \qquad\qquad \ov D^\dalpha J_{\alpha\dalpha} = D_\alpha X\ne0.
\eeq
Since $X\ne0$, the $U(1)_R$ current $j_\mu$ is not conserved and the trace of the
energy-momentum tensor in $J_{\alpha\dalpha}$ is correlated by supersymmetry with $\partial^\mu j_\mu$,
see eqs.~(\ref{4sc10}).
Since $\chi_\alpha = - {1\over4}\ov{DD}D_\alpha U$ in a generic supercurrent stucture, it can be in principle 
eliminated using the superfield improvement transformation (\ref{4sc12}) with ${\cal G} = {1\over12}U$, 
to obtain a FZ structure. Problems could arise if for instance $U$ would not respect symmetries of the 
underlying theory. The simplest example would be a symmetry acting on $U$ with 
$\delta U = {\cal F} + \ov{\cal F}$, where ${\cal F}$ is a chiral function 
(leaving $\chi_\alpha$ unchanged).\footnote{Theories with Fayet-Iliopoulos terms are not problematic 
\cite{ADH}.}
The FZ structure is not unique: it is preserved by improvement transformations (\ref{4sc12}) with 
${\cal G}=\Psi+\ov\Psi$, $\ov D_\dalpha\Psi=0$. 

Secondly, the {\it $R$--invariant structure} with $12_B+12_F$ component fields or operators:
\beq
\makebox{$R$--invariant structure:} \qquad\qquad
X=0, \qquad\qquad \ov D^\dalpha J_{\alpha\dalpha} = \chi_\alpha\ne 0 .
\eeq 
Since $X=0$, the supercurrent superfield $J_{\alpha\dalpha}$ includes the current $j_\mu$ of an exact 
$R$--symmetry in its lowest component and the traces ${T^\mu}_\mu$ and $(\ov\sigma^\mu S_\mu)^\dalpha$ are
not zero in general. 
This structure can be obtained whenever $X=\ov{DD}\,{\cal U}$ for some real superfield ${\cal U}$. In this
case $\Im f_X$ is itself the divergence of a vector field $V_\mu$ (off-shell)\footnote{The condition that a vector field
$V_\mu$ exists with $\Im f_X=\partial^\mu V_\mu$ is equivalent 
to the existence condition of ${\cal U}$, with $X=\ov{DD}\,{\cal U}$.} 
and $\partial^\mu (j_\mu - V_\mu)=0$. The source superfield $X$ can then be eliminated by the superfield 
improvement (\ref{4sc12}) with ${\cal G}= -{\cal U}$.
The transformed $J_{\alpha\dalpha}$ has lowest component $j_\mu - V_\mu$. An obstruction can exist if 
${\cal U}$ is not invariant under symmetries of the underlying symmetry.
The $R$--invariant structure is preserved by improvement transformations (\ref{4sc12}) with 
${\cal G}$ real linear ($\ov{DD}\,{\cal G}=0$). 

Thirdly, the {\it superconformal} structure with $8_B+8_F$ component fields or operators:
\beq
\makebox{Superconformal structure:} \qquad\qquad
X=\chi_\alpha=0, \qquad\qquad \ov D^\dalpha J_{\alpha\dalpha}=0.
\eeq
It can be obtained whenever the source superfields in a generic structure are generated by a single real 
superfield ${\cal G}$: $X=\ov{DD}\,{\cal G}$ and $\chi_\alpha=3\,\ov{DD}D_\alpha{\cal G}$. 
A superfield improvement (\ref{4sc12}) leads then to a superconformal structure with $X=\chi_\alpha=0$.
In this case, the theory admits a conserved, symmetric and traceless symmetric energy-momentum tensor: it 
is conformal. In addition, it has an exact $R$--symmetry and a conserved supercurrent $S_{\mu\alpha}$ with 
zero $\gamma$-trace, $(\ov\sigma^\mu S_\mu)^\dalpha=0$: the theory is superconformal.
If the supersymmetric theory is coupled to conformal ${\cal N}=1$ supergravity, the conserved currents 
$j_\mu$, $T_{\mu\nu}$ and $S_{\mu\alpha}$ ($8_B+8_F$) couple to gauge fields of the superconformal algebra 
\beq
T_{\mu\nu} \quad\longleftrightarrow\quad g_{\mu\nu}, \qquad\quad
S_{\mu\alpha} \quad\longleftrightarrow\quad \psi_{\mu\alpha}, \qquad\quad
j_\mu \quad\longleftrightarrow\quad A_\mu ,
\eeq
where $\psi_{\mu\alpha}$ is the gravitino and $A_\mu$ the $U(1)_R$ gauge field.

A theory with a FZ or a $R$--invariant supercurrent structure is not superconformal. It couples to Poincar\'e 
supergravity which can be obtained by gauge-fixing a superconformal theory, using various sets of 
compensating fields: this procedure leads to various formulations of Poincar\'e supergravity 
characterized by their auxiliary field content \cite{superconf, KU, FGKVP}. 

The chiral source multiplet $X$ of the FZ structure corresponds to the chiral compensating multiplet $S_0$ 
(with nonzero Weyl weight, usually $w=1$ and $R$--charge $q=w$)
used in {\it old minimal}\, supergravity \cite{oldmin}, with $12_B+12_F$ component fields in the off-shell Poincar\'e supergravity multiplet (and auxiliary fields $A_\mu$, with $4_B$ fields, and a complex scalar ($2_B$) $f_0$). 

The source supermultiplet $\chi_\alpha$ of the $R$--invariant structure naturally couples to the real linear compensating multiplet $L_0$ ($w=2$) used in {\it new minimal}\, supergravity \cite{newmin}, with $12_B+12_F$ fields
in the off-shell Poincar\'e supermultiplet. The auxiliary fields are an antisymmetric tensor $B_{\mu\nu}$ with 
gauge invariance ($3_B$) and the gauge field $A_\mu$ ($3_B$).

The generic structure with $X\ne0\ne\chi_\alpha$ and $16_B+16_F$ components couples finally to a conformal 
supergravity with both chiral and linear supermultiplets with nonzero Weyl weight. These multiplets provide the compensating fields for Poincar\'e gauge-fixing, supergravity auxiliary fields and $4_B+4_F$
propagating fields of the globally supersymmetric theory. 

\section{Supercurrent structures of supersymmetric \\ gauge theories} \label{secsusygauge}

In general, the construction of currents and of their (non-)conservation equations begins with an identity
which, in essence, does not carry information. It acquires significance when field equations of a
given theory are applied. 
In the following, we apply this method to derive supercurrent structures of generic (two-derivative) ${\cal N}=1$ 
supersymmetric theories.\footnote{This section mostly follows refs.~\cite{ADH, AADH}.}

\subsection{Identities}

This subsection is purely technical. We use the following superfields:
\begin{itemize}
\item
A set of chiral superfields $\Phi$, $\ov D_\dalpha\Phi=0$ and their conjugate antichiral superfields
$\ov\Phi$, $D_\alpha\ov\Phi=0$. These fields are in a representation $r$, in general reducible, of the 
gauge group.\footnote{Component fields: complex scalars $z$, Weyl spinors $\psi$, complex auxiliary scalars $f$.}
\item
Gauge superfields: the real superfield of gauge fields ${\cal A}$ and the chiral superfield of
gauge curvatures (field strengths)\footnote{Component fields in Wess-Zumino gauge: 
gauge fields ${\cal A}_\mu$ with field strengths
$F_{\mu\nu}$, gauginos $\lambda$, real auxiliary scalars $D$.}
\beq
{\cal W}_\alpha({\cal A}) = -{1\over4}\ov{DD} \, e^{-{\cal A}}D_\alpha e^{\cal A},
\qquad\qquad
\ov {\cal W}_\dalpha = \frac{1}{4}DD\, e^{\cal A}\ov D_\dalpha e^{-{\cal A}}. 
\eeq
They are Lie algebra-valued, with ${\cal A} = {\cal A}^aT_r^a$ and generators $T_r^a$ for representation $r$, 
normalized with $\Tr(T_r^aT_r^b) = T(r)\delta^{ab}$.
We will also use the real Chern-Simons superfield $\Omega$ defined by
\beq
\label{lindef}
\ov{DD} \,\Omega = \widetilde\Tr{\cal WW}, \qquad\qquad DD\, \Omega = \widetilde\Tr\ov{\cal WW},
\eeq
using the notation
$$
\widetilde\Tr {\cal WW} = T(r)^{-1} \Tr {\cal WW}.
$$
The gauge variation of $\Omega$ is linear, $\ov{DD}\,\delta\Omega=DD\,\delta\Omega=0$. Closed expressions
for $\Omega$ are easily obtained in the abelian case:
\beq
\begin{array}{rcl}
\Omega &=&\displaystyle 
-{1\over 4} \Bigl[ {\cal W}^\alpha D_\alpha{\cal A} + \ov D_\dalpha[ {\cal A}\ov{\cal W}^\dalpha] \Bigr]
= -{1\over 4} \Bigl[ \ov{\cal W}_\dalpha\ov D^\dalpha{\cal A} + D^\alpha[ {\cal A}{\cal W}_\alpha]\Bigr]
\crbig
&=& \displaystyle -{1\over 4} \Bigl[ {\cal W}^\alpha D_\alpha{\cal A} + \ov{\cal W}_\dalpha\ov D^\dalpha{\cal A}
+ {1\over2}{\cal A} [ D^\alpha{\cal W}_\alpha + \ov D_\dalpha\ov{\cal W}^\dalpha ] \Bigr].\end{array}
\eeq
The first two expresssions manifestly verify one of the two conditions (\ref{lindef}), the third expression is 
manifestly hermitian, the equalities follow from the abelian Bianchi identity $D^\alpha{\cal W}_\alpha = 
\ov D_\dalpha\ov{\cal W}^\dalpha$. The non-abelian $\Omega$ is much more subtle \cite{CFV}.
\item
A linear superfield $L$,\,\footnote{Component fields: real scalar $C$, antisymmetric
tensor $B_{\mu\nu}$ in the gauge-invariant curl $H_{\mu\nu\rho} = 3\,\partial_{[\mu}B_{\nu\rho]}$, spinor $\chi$.}
which will be used as the gauge coupling superfield. It is real with
$\ov{DD}\,L=DD\,L=0$, hence the terminology {\it linear}. It will be coupled to the Chern-Simons superfield
to form the gauge-invariant and real 
\beq
\hat L = L-2\,\Omega
\eeq
with the postulate that the gauge variations of $L$ and $\Omega$ cancel in $\hat L$: 
$\delta L=2\,\delta\Omega$.\footnote{
If the gauge group has several simple or $U(1)$ factors, we could introduce one Chern-Simons superfield 
$\Omega_i$ and one linear superfield $L_i$ for each factor, or define a gauge-invariant
$\hat L = L - 2\sum_i c_i\Omega_i$. } 
\end{itemize}
The first identity applies to an arbitrary real function ${\cal H}$ of the gauge-invariant superfields
$\hat L$ and
\beq
Y = \ov\Phi e^{\cal A}\Phi.
\eeq
By direct calculation of, for instance,  $\ov{DD}D_\alpha({\cal H}-\hat L{\cal H}_L)$, one 
obtains\footnote{In general, the gauge invariant function ${\cal H}$ can depend on
variables $Y_i$ if the representation of the chiral superfields is reducible, $r=\oplus_ir_i$. This
generalization is straightforward. It may also depend on other gauge invariant quantities,
such as holomorphic invariants, which we do not consider here.}
\beq
\label{Id1}
\begin{array}{l}
\makebox{Id\,1:}\qquad 2 \ov D^{\dalpha}\Bigl[ (\ov{\cal D}_\dalpha\ov\Phi) {\cal H}_{\Phi\ov\Phi}  ({\cal D}_\alpha\Phi)
- {\cal H}_{LL}(\ov D_\dalpha\hat L)(D_\alpha\hat L) \Bigr]
\crbig \hspace{2.5cm}
= - \hat L \, \ov{DD}D_\alpha{\cal H}_L - (\ov{DD}{\cal H}_\Phi)\, {\cal D}_\alpha\Phi
- \ov{DD}D_\alpha({\cal H}-\hat L{\cal H}_L)
\crbig\hspace{2.9cm}
- 2 \,  \widetilde\Tr\,{\cal WW} \, D_\alpha{\cal H}_L - 4 \, {\cal H}_Y \, \ov\Phi e^{A}{\cal W}_\alpha\Phi ,
\end{array}
\eeq
where subscripts indicate 
derivatives of ${\cal H}$ with respect to either $\Phi$, $\ov\Phi$, $\hat L$ or $Y$. 
Gauge transformations are
\beq
\label{CSF1}
\begin{array}{c}
\Phi\quad\longrightarrow\quad e^\Lambda\, \Phi, \qquad
\ov\Phi\quad\longrightarrow\quad \ov\Phi \,e^{\ov\Lambda}, \qquad
e^{\cal A} \quad\longrightarrow\quad e^{-\ov\Lambda} e^{\cal A} e^{-\Lambda} ,
\crbig
{\cal W}_\alpha \quad\longrightarrow\quad e^\Lambda {\cal W}_\alpha e^{-\Lambda}, 
\qquad\qquad
\ov {\cal W}_\dalpha \quad\longrightarrow\quad e^{-\ov\Lambda}\ov {\cal W}_\dalpha e^{\ov\Lambda},
\end{array}
\eeq
with $\Lambda = \Lambda^aT^a_r$ and $\ov D_{\dot\alpha}\Lambda=0$. Gauge-covariant superspace 
derivatives read
\beq
\label{CSF2}
{\cal D}_\alpha \Phi = e^{-{\cal A}}(D_\alpha e^{\cal A}\Phi) , 
\qquad\qquad
\ov{\cal D}_\dalpha \ov\Phi = (\ov D_\dalpha\ov\Phi e^{\cal A}) e^{-{\cal A}}
\eeq
and 
$$
(\ov{\cal D}_\dalpha\ov\Phi)e^{\cal A}({\cal D}_\alpha\Phi)
= (\ov D_\dalpha\ov\Phi e^{\cal A}) e^{-{\cal A}} (D_\alpha e^{\cal A}\Phi)
$$ 
is gauge invariant. 
Removing the linear superfield  with ${\cal H}_L=0$ leads to
\beq
\label{Id1b}
2\,\ov D^\dalpha \Bigl[(\ov{\cal D}_\dalpha \ov\Phi)K_{\Phi\ov\Phi}
({\cal D}_\alpha\Phi)\Bigr]
= -\ov{DD}D_\alpha K - 4 K_\Phi {\cal W}_\alpha\Phi - (\ov{DD} K_\Phi) ({\cal D}_\alpha\Phi).
\eeq 
for an arbitrary function ${\cal K}(\Phi, \ov\Phi e^{\cal A})$.
Gauge invariance reads $ {\cal K}_\Phi T^a_r \Phi = \ov\Phi T^a_r {\cal K}_{\ov\Phi}$ for all generators.

We also need identities for gauge superfields. The tool is the 
non-abelian Bianchi identity:
\beq
\label{WZ10}
e^{-{\cal A}}D^\alpha(e^{\cal A} {\cal W}_\alpha e^{-{\cal A}}) e^{\cal A} 
= \ov D_\dalpha(e^{-{\cal A}}\ov{\cal W}^\dalpha e^{\cal A}).
\eeq
Multiplying (left) by ${\cal W}_\alpha$ and taking the trace gives
\beq
\label{Id2}
\makebox{Id\,2:}\qquad
\ov D^\dalpha \widetilde\Tr [ {\cal W}_\alpha\, e^{-{\cal A}}\ov {\cal W}_\dalpha e^{\cal A}  ]
= \widetilde\Tr [ e^{\cal A} {\cal W}_\alpha e^{-{\cal A}} D^\beta ( e^{\cal A}{\cal W}_\beta e^{-{\cal A}} ) ].
\eeq
Then, for an arbitrary (gauge-invariant) holomorphic function $F(\Phi)$,
\beq
\label{Id3}
\begin{array}{rcl}
\makebox{Id\,3:}\hspace{7mm}
\ov D^\dalpha \Bigl[ (F+\ov F)\widetilde\Tr [ {\cal W}_\alpha\, e^{-{\cal A}}\ov {\cal W}_\dalpha e^{\cal A} ] \Bigr]
&=& (F+\ov F)\, \widetilde\Tr [ e^{\cal A}{\cal W}_\alpha e^{-{\cal A}} D^\beta ( e^{\cal A}{\cal W}_\beta e^{-{\cal A}} ) ]
\crbig
&& + (\ov D^\dalpha\ov F)  \widetilde\Tr [ {\cal W}_\alpha\, e^{-{\cal A}}\ov {\cal W}_\dalpha e^{\cal A}  ].
\end{array}
\eeq
For given superspace lagrangians and the corresponding superfield dynamical equations, these identities
``automatically" produce supercurrent structures.

\subsection{The natural supercurrent structure}

Let us  consider theory
\beq
\label{CSF4}
\L=\Dint {\cal H} ( \hat L , Y ) + \Fint W(\Phi) + \Fbarint \ov W(\ov\Phi).
\eeq
Gauge invariance of the holomorphic superpotential $W(\Phi)$, {\it i.e.}\,\,$W_{\Phi^i}(T_r^a)^i{}_j\Phi^j=0$,
implies $W_\Phi{\cal D}_\alpha\Phi  = D_\alpha W$. The ${\cal H}$ term in the lagrangian has in general several 
chiral symmetries. In particular, since $\cal H$ satisfies
\beq
\label{CSF4b}
{\cal H}_\Phi\Phi=\ov\Phi {\cal H}_{\ov\Phi}={\cal H}_YY,
\eeq
it is always invariant under the non-$R$ $U(1)$ symmetry 
rotating all chiral superfields $\Phi$ by the same phase.\footnote{If the representation of the 
matter superfields is reducible, each irreducible component has an associated $U(1)$ global 
symmetry. It extends to $U(n)$ factors if the matter superfields include $n$ 
copies of an irreducible component.} Its chiral symmetries also 
include the $R$ symmetry (that we call $\widetilde R$) which transforms Grassmann coordinates and leaves 
superfields $\hat L$ and $\Phi$ inert. These chiral symmetries are in general broken 
by the superpotential. 

The component expansion of theory (\ref{CSF4}) is\footnote{Gauge 
invariance of ${\cal H}$ implies ${\cal H}_z Dz = \ov z D{\cal H}_{\ov z}$. }
\beq
\label{bosonicL}
\begin{array}{rcl}
{\cal L} &=& -{1\over2}{\cal H}_{CC} \Bigl[ {1\over2}(\partial_\mu C)(\partial^\mu C) 
+ {1\over12} H_{\mu\nu\rho}H^{\nu\mu\rho}  \Bigr]
+ {\cal H}_{z\ov z} \Bigl[ (D_\mu \ov z)(D^\mu z) + \ov ff \Bigr] 
\crbig
&& + {\cal H}_C \Bigl[ -{1\over4}\widetilde\Tr F_{\mu\nu}F^{\mu\nu} + {1\over2}\widetilde\Tr DD \Bigr]
+ {1\over2}{\cal H}_z D z - W_zf - \ov f \ov W_{\ov z}
\crbig
&& + {i\over12}\epsilon_{\mu\nu\rho\sigma}
H^{\mu\nu\rho} \Bigr[ {\cal H}_{Cz}D^\sigma z - {\cal H}_{C\ov z}D^\sigma\ov z \Bigr]
\makebox{$+$ fermion terms}\,,
\end{array}
\eeq
with covariant derivative $(D_\mu z)^i = \partial_\mu z^i+ {i\over2} A^a_\mu(T_r^a)^i{}_jz^j$ and with
\beq
H_{\mu\nu\rho} = h_{\mu\nu\rho}-\omega_{\mu\nu\rho}\,,
\eeq
in terms of the Chern--Simons form $\omega$ with normalization such that $dH=-\widetilde\Tr F\wedge F$. 
The kinetic metrics are then ${\cal H}_{z\ov z}$, $-{1\over2}{\cal H}_{CC}$ and ${\cal H}_C$ for the components of superfields $\Phi$, $L$ and ${\cal W}_\alpha$ respectively.

The field equations for theory (\ref{CSF4}) are\footnote{We use the convention 
$\ov {\cal W}_\dalpha = {1\over4}DD e^{\cal A} \ov D_\dalpha e^{-{\cal A}}$, with 
$\ov{\cal W}_\dalpha = - ({\cal W}_\alpha)^\dagger$. }
\beq
\label{CSF5}
\begin{array}{rrcl}
L:& \qquad\qquad \ov{DD}D_\alpha{\cal H}_L &=& 0,
\crbig
\Phi:& \ov{DD}{\cal H}_\Phi &=& 4\,W_\Phi,
\crbig
{\cal A} : &  \ov D^\dalpha \Bigl[ {\cal H}_L \, e^{-{\cal A}} \ov {\cal W}_\dalpha e^{\cal A} \Bigr] 
&=& {\cal W}^\alpha \, D_\alpha{\cal H}_L - T(r)\,{\cal H}_Y\, \Phi \ov\Phi e^{\cal A},
\end{array}
\eeq
with index $\Tr(T^a_rT^b_r) = T(r)\delta^{ab}$.
To derive the field equation for the gauge superfield ${\cal A}$, it is indeed easier to use the dual chiral 
version of the theory,\footnote{To avoid dealing with the complicated non-Abelian 
Chern-Simons superfield \cite{CFV}.}
\beq
\label{CSF6}
\begin{array}{rcl}
{\cal L} &=& \Dint {\cal K}(S+\ov S, Y)
\crbig
&&+ \Fint \left[ W(\Phi) + {1\over4}S \,\widetilde \Tr\,{\cal WW}\right] 
+ \Fbarint \left[ \ov W(\ov\Phi) + {1\over4} \ov{S}\,\widetilde\Tr\,\ov{\cal WW} \right] ,
\end{array}
\eeq
where ${\cal K}$ is the Legendre transform of ${\cal H}$, 
and to transform the resulting field equation back into the linear version. 
Variation of eq.~(\ref{CSF6}) and use of the Bianchi identity (\ref{WZ10}) 
gives then the field equation
\beq
\label{CSF9}
\ov D^\dalpha \Bigl[ (S+\ov S) \, e^{-{\cal A}} \ov {\cal W}_\dalpha e^{\cal A} \Bigr] = 
D^\alpha(S+\ov S) \, {\cal W}_\alpha
- 2\,T(r)\,{\cal K}_Y\, \Phi \ov\Phi e^{\cal A} .
\eeq
Multiplying by ${\cal W}_\beta$ and taking the trace gives
\beq
\label{CSF10}
\ov D^\dalpha \Bigl[ (S+\ov S) \Tr ({\cal W}_\beta e^{-{\cal A}} \ov {\cal W}_\dalpha e^{\cal A}) \Bigr] 
= {1\over2} D_\beta (S+\ov S) \, \Tr{\cal WW} 
+ 2\,T(r)\,{\cal K}_Y\,  \ov\Phi e^{\cal A} {\cal W}_\beta \Phi.
\eeq
The Legendre transformation indicates then that ${\cal K}_Y = {\cal H}_Y$ and 
$S+\ov S = 2{\cal H}_L$, which in turn implies the field equation (\ref{CSF5}) for ${\cal A}$ and the relation
\beq
\label{CSF11}
\ov D^\dalpha \Bigl[ {\cal H}_L \Tr ({\cal W}_\beta e^{-{\cal A}} \ov {\cal W}_\dalpha e^{\cal A}) \Bigr] 
= {1\over2} D_\beta {\cal H}_L \, \Tr{\cal WW} 
+ T(r){\cal H}_Y\,  \ov\Phi e^{\cal A} {\cal W}_\beta \Phi.
\eeq
With field equations (\ref{CSF5}) and relation (\ref{CSF11}), identity (\ref{Id1}) immediately leads to the 
supercurrent structure
\beq
\label{CSF12}
\begin{array}{rcl}
\ov D^\dalpha J_{\alpha\dalpha} &=& D_\alpha X + \chi_\alpha ,
\crbig
J_{\alpha\dalpha} &=&
- 2 \Bigl[ (\ov{\cal D}_\dalpha\ov\Phi) {\cal H}_{\Phi\ov\Phi}  ({\cal D}_\alpha\Phi)
- {\cal H}_{LL}(\ov D_\dalpha\hat L)(D_\alpha\hat L) 
+ 2 \, {\cal H}_L \widetilde\Tr ({\cal W}_\alpha e^{-{\cal A}} \ov {\cal W}_\dalpha e^{\cal A}) \Bigr] ,
\crbig
X &=& 4\, W, 
\crbig
\chi_\alpha &=& \ov{DD}D_\alpha({\cal H}-\hat L{\cal H}_L) .
\end{array}
\eeq
This supercurrent structure can be considered as {\it natural} for theory (\ref{CSF4}). 
It actually also applies if ${\cal H}$ is simply a gauge-invariant function of $\hat L$, $\Phi$ and 
$\ov\Phi e^{\cal A}$, instead of a function of $\hat L$ and $Y$.

Using expansion (\ref{4sc8}) of the superfield $J_\mu=(\ov\sigma_\mu)^{\dalpha\alpha}J_{\alpha\dalpha}$
and also
$$
\hat L = C + i\theta\chi - i\ov\theta\ov\chi + \ldots,
\qquad
\Phi = z + \sqrt2\,\theta\psi-\theta\theta f+\ldots ,
\qquad
{\cal W}_\alpha = -i\lambda_\alpha + \ldots ,
$$
the lowest component of the supercurrent superfield (\ref{CSF12}) is 
\beq
\label{CSF13}
j^{\widetilde R}_\mu\equiv  {3\over8}(\ov\sigma_\mu)^{\dalpha\alpha}J_{\alpha\dalpha}\big|_{\theta=0}=-{3\over2}\,{\cal H}_{z\ov z} \,\psi \sigma_\mu\ov\psi + {3\over4}\,{\cal H}_{CC}\, \chi\sigma_\mu\ov\chi 
+ {3\over2}\,{\cal H}_C\, \widetilde\Tr \lambda\sigma_\mu\ov\lambda \, .
\eeq
It is the Noether current of 
$\widetilde R$--transformations with chiral charges $-3/2$, $-3/2$ and $3/2$ for $\chi$, $\psi$ 
and $\lambda$ 
respectively. The chiral charges of superfields $\Phi$, $L$ and ${\cal W}_\alpha$ 
for this $U(1)_{\widetilde R}$ are then $q=0$, $0$, $3/2$ in this supercurrent structure and
$U(1)_{\widetilde R}$ only acts on the Grassmann coordinates.\footnote{The charge $q=3/2$ 
of ${\cal W}_\alpha$
is due to the derivatives in ${\cal W}_\alpha = -{1\over4}\ov{DD} e^{-{\cal A}} D_\alpha e^{\cal A}$.}
It is an automatic symmetry of
$D$--term lagrangians and, according to the second eq.~(\ref{4sc10}), 
the $\widetilde R$ current is conserved if the superpotential vanishes,
$\partial^\mu j_\mu^{\widetilde R} = -{3\over2}\Im f_X$. 

The supercurrent superfield $J_{\alpha\dalpha}$ of eqs.~(\ref{CSF12}) also contains the Belinfante 
(symmetric, gauge-invariant) energy-momentum tensor $T_{\mu\nu}$ for theory (\ref{CSF4}). 
Omitting fermions and gauge fields, its expression is 
\beq
\label{CSF13b}
\begin{array}{rcl}
T_{\mu\nu}Ê&=& -{1\over2} {\cal H}_{CC} (\partial_\mu C)(\partial_\nu C)
- {1\over4} {\cal H}_{CC}h_{\mu\rho\sigma}{h_\nu}^{\rho\sigma}
+  {\cal H}_{z\ov z}[(\partial_\mu z)(\partial_\nu\ov z) + (\partial_\nu z)(\partial_\mu\ov z)]
\crbig
&& - \eta_{\mu\nu} \Bigl(  -{1\over4} {\cal H}_{CC} (\partial_\rho C)(\partial^\rho C)
- {1\over24} {\cal H}_{CC}h_{\rho\sigma\lambda}h^{\rho\sigma\lambda}
+ {\cal H}_{z \ov z} [ (\partial_\rho z)(\partial^\rho\ov z) + \ov ff ]  \Bigr)
\crbig
&& + {1\over2} \eta_{\mu\nu} {\cal H}_C \widetilde\Tr(D^2) + {1\over2}\eta_{\mu\nu} \Re f_X ,
\end{array}
\eeq
with auxiliary fields\footnote{The auxiliary field contribution to $T_{\mu\nu}$ is
$\eta_{\mu\nu}V$, where $V$ is the usual scalar potential 
$$
V (C, z, \ov z) = {1\over2} {\cal H}_C \widetilde\Tr D^2 + {\cal H}_{\ov zz}\ov f f.
$$} 
$$f_X = 4W_zf, \qquad \ov f {\cal H}_{\ov zz} = W_z, \qquad
D^a = -{1\over2} {\cal H}_C^{-1} {\cal H}_z T^a_r z
= -{1\over2} {\cal H}_C^{-1} {\cal H}_Y \ov z T^a_rz.
$$
Notice that terms depending on 
${\cal H}_{Cz}$ or ${\cal H}_{C\ov z}$ present in the lagrangian do not appear
in the Belinfante tensor $T_{\mu\nu}$. If the superpotential vanishes, $f=f_X=0$.

Hence, the Belinfante tensor and the $\widetilde R$ current with zero charge chiral superfields are partners
in the natural supercurrent structure.
 
\subsection{The improved supercurrent structure}

Suppose that we assign scale dimensions $w$ and $R$--charges $q$ to the chiral superfields $\Phi$.\footnote{We 
suppress indices.
The introduction of independent $w_i$ and $q_i$ for each irreducible component $\Phi_i$ is straightforward. 
The scale dimension of $\hat L$ is always $w=2$: the dimension of $\Omega$ is canonical. 
The linear $L$ contains a dimension-three vector field 
$\epsilon_{\mu\nu\rho\sigma}\partial^\nu b^{\rho\sigma}$ which is transverse, $\partial^\mu v_\mu=0$.} 
The behaviour of theory (\ref{CSF4}) under dilatations is controlled by two superfields:
\beq
\label{Deltais}
\begin{array}{rcll}
\Delta_{(w)} &=& w\Phi{\cal H}_\Phi + w\ov\Phi{\cal H}_{\ov\Phi} + 2\hat L{\cal H}_{\hat L} - 2 {\cal H}
\qquad
&\makebox{(real)},
\crbig
\widetilde\Delta_{(w)} &=& w\Phi W_\Phi - 3 W
&\makebox{(chiral)}.
\end{array}
\eeq
Scale invariance is obtained if $\Delta_{(w)}=\widetilde\Delta_{(w)}=0$ for some $w$.
Similarly, the variation under $R$ is controlled by
\beq
\label{Xiis}
\begin{array}{rcll}
\Xi_{(q)} &=& i q ( \Phi{\cal H}_\Phi - \ov\Phi{\cal H}_{\ov\Phi}) \qquad
&\makebox{(real)},
\crbig
\widetilde\Xi_{(q)} &=& q\Phi W_\Phi - 3 W \,\,=\,\, \widetilde\Delta_{(q)} \qquad
&\makebox{(chiral)}.
\end{array}
\eeq
The $R$--current (\ref{CSF13}) indicates that chiral superfields in the natural structure (\ref{CSF12}) have 
zero charge, $\Xi_{(0)}=0$, and the source superfields of this structure are then
\beq
X = - {4\over3} \, \widetilde\Delta_{(0)} = - {4\over3} \, \widetilde\Xi_{(0)} ,
\qquad\qquad
\chi_\alpha = - {1\over2} \ov{DD}D_\alpha \, \Delta_{(0)}.
\eeq
From theory (\ref{CSF4}), one easily deduces the dilatation current, expressed in terms of the Belinfante
tensor, and its divergence (using field equations). There is of course a virial current for the scalar fields, 
\beq
\label{CSF13f}
j_\mu^{D} = -{1\over2}\left[ {\partial\over\partial C}\Delta_{(0)} \Bigr|_{\theta=0} \,\partial_\mu C \right] 
+ x^\nu T_{\mu\nu} ,
\eeq
and it is not a derivative in general: the linear superfield coupled to chiral fields opposes the
existence of the CCJ tensor. But the virial current also cancels with scale invariance condition $\Delta_{(0)}=0$.
For the natural structure, $R$--charges and scale dimensions of $\Phi$ vanish.
We now wish to obtain supercurrent structures for nonzero weights of $\Phi$. 

Applying to the natural structure (\ref{CSF12}) the superfield improvement transformation (\ref{4sc12}) 
with
\beq
\label{Imp3}
{\cal G} = -{w\over6}({\cal H}_\Phi\Phi + \ov\Phi{\cal H}_{\ov\Phi}),
\eeq
the chiral source superfield $X$ becomes
\beq
\label{Imp5}
\widetilde X = -{4\over3} \widetilde\Delta_{(w)} + {4\over3}w \, W_\Phi\Phi
-{w\over6}\ov{DD}({\cal H}_\Phi\Phi + \ov\Phi{\cal H}_{\ov\Phi}) 
\eeq
or
\beq
\widetilde X = -{4\over3} \widetilde\Delta_{(w)} + {w\over6}\ov{DD}({\cal H}_\Phi\Phi - \ov\Phi{\cal H}_{\ov\Phi})
= -{4\over3} \widetilde\Delta_{(w)} - {i\over6}\ov{DD}\,\Xi_{(w)}
\eeq
using the field equation of $\Phi$. The resulting {\it improved supercurrent structure} is then 
\beq
\label{Imp4}
\begin{array}{rcl}
\ov D^\dalpha \widetilde J_{\alpha\dalpha} &=& D_\alpha \widetilde X + \widetilde\chi_\alpha ,
\crbig
\widetilde J_{\alpha\dalpha} &=&
- 2 \Bigl[ (\ov{\cal D}_\dalpha\ov\Phi) {\cal H}_{\Phi\ov\Phi}  ({\cal D}_\alpha\Phi)
- {\cal H}_{LL}(\ov D_\dalpha\hat L)(D_\alpha\hat L) 
+ 2 \, {\cal H}_L \widetilde\Tr ({\cal W}_\alpha e^{-{\cal A}} \ov {\cal W}_\dalpha e^{\cal A}) \Bigr] 
\crbig
&& \hspace{2.2cm}
-{w\over3} [D_\alpha,\ov D_\dalpha] ({\cal H}_\Phi\Phi + \ov\Phi{\cal H}_{\ov\Phi}) ,
\crbig
\widetilde X &=& -{4\over3} \widetilde\Delta_{(w)} + {w\over6}\ov{DD}({\cal H}_\Phi\Phi - \ov\Phi{\cal H}_{\ov\Phi}),
\crbig
\widetilde \chi_\alpha &=& -{1\over2}\ov{DD}D_\alpha\Delta_{(w)} .
\end{array}
\eeq
In the canonical Wess-Zumino model, ${\cal H} = \ov\Phi\Phi$, the supercurrent superfield reduces to 
\beq
\widetilde J_{\alpha\dalpha} = {4\over3}\left[ \left( w-{3\over2}\right) (\ov D_\dalpha\ov\Phi)(D_\alpha\Phi) 
-iw \, (\sigma^\mu)_{\alpha\dalpha}\, \ov \Phi \stackrel{\leftrightarrow}{\partial_\mu} \Phi \right]
\eeq 
with $R$--current
\beq
j_\mu = \left( w-{3\over2}\right)  \psi\sigma_\mu\ov\psi - iw \ov z \stackrel{\leftrightarrow}{\partial_\mu} z ,
\eeq
two results often used in the literature with canonical scale dimension or $R$--charge $w=1$. 

As required for superconformal invariance, the 
source superfields in structure (\ref{Imp4}) vanish if two conditions are fulfilled. Firstly, scale invariance
$\Delta_{(w)}=\widetilde\Delta_{(w)}=0$ and secondly that the theory has a $U(1)$ $R$--symmetry rotating
$\Phi$ with charges $q=w$: $w ({\cal H}_\Phi\Phi - \ov\Phi{\cal H}_{\ov\Phi})=0$. This second condition 
is certainly fulfilled if ${\cal H}$ is a fonction of $\hat L$ and $Y$. If it is not verified, scale invariance may not 
imply conformal invariance. A simple example is the K\"ahler potential $K={1\over2}(\Phi^2\ov\Phi+\ov\Phi^2\Phi)$ 
for a single chiral superfield: the CCJ energy-momentum tensor does not exist and scale invariance with 
$w=2/3$ does not imply conformal invariance.
In $\widehat J_{\alpha\dalpha}$, the energy-momentum tensor 
$\Theta_{\mu\nu}$ is related to the Belinfante tensor by the improvement
\beq
\label{Imp7}
\begin{array}{rcl}
\Theta_{\mu\nu} &=& T_{\mu\nu} - {1\over6} (\partial_\mu\partial_\nu - \eta_{\mu\nu}\Box)
w({\cal H}_zz + \ov z{\cal H}_{\ov z}) \crbig
&=& T_{\mu\nu} - {1\over3} (\partial_\mu\partial_\nu - \eta_{\mu\nu}\Box)
w{\cal H}_yy , \qquad\qquad y=\ov zz.
\end{array}
\eeq
The virial current derived from the difference between the divergence of the dilatation current, which is not in
the supercurrent structure, and ${\Theta^\mu}_\mu$ is
\beq
\label{Imp8}
\widehat{\cal V}_\mu = - {1\over2} {\partial\Delta_{(w)}\over\partial C} \,\partial_\mu C. 
\eeq
It vanishes if $\Delta_{(w)}=0$, {\it i.e.} if ${\cal H}$ has scale dimension two for scale dimensions $w$. 
Requiring the existence of the CCJ energy-momentum tensor selects a particular class of functions ${\cal H}$
where $\partial_{\hat L}\Delta_{(w)}$ is a function of $\hat L$ only, for a choice of $w$:
\beq
\label{CL3a}
{\cal H}(\hat L,Y) = {\cal F}_1(\hat L) + {\cal F}_2(Y) + {\cal I}(\hat L,Y), 
\qquad\qquad
wY{\cal I}_Y + \hat L{\cal I}_L = {\cal I}.
\eeq
The second equation indicates that the coupling of chiral to matter multiplets should have scale dimension
two, and the corresponding interaction terms should be scale invariant. For instance, 
\beq
{\cal I}(\hat L, Y) = \hat L \, \widetilde {\cal I}(X), \qquad\qquad
X = Y \hat L^{-w}
\eeq
allows to find an energy-momentum tensor such that 
$\partial^\mu j_\mu^{D} = {\Theta^\mu}_\mu$.

The supercurrent superfield $\widehat J_{\alpha\dalpha}$ includes in its lowest component the 
current of the $R$ transformation with $R$ charges $0$ and $w$ for $\hat L$ and $\Phi$ respectively. Gauginos,  
fermions $\psi$ in $\Phi$ and $\chi$ in $L$ have chiral weights $3/2$, $w-3/2$ and $-3/2$ respectively.
Notice that $w$ has been originally introduced as the scale dimension 
of $\Phi$ and it here also plays the role of an $R$ charge. This is reminiscent of the chirality condition
in a superconformal theory, in which the scale dimension and the $U(1)_R$ charge are identified.
This $R$ transformation combines $\widetilde R$ and a $U(1)_{\cal Z}$ non--$R$ transformation acting
on $\Phi$ with charge $w$. The non-conservation equation for the vector superfield
${\cal Z}$ of $U(1)_{\cal Z}$ is
\beq
\ov{DD}{\cal Z} = 4wW_\Phi\Phi
- {1\over2}\ov{DD}(w{\cal H}_\Phi\Phi - w\ov\Phi{\cal H}_{\ov\Phi}),
\qquad\quad
{\cal Z} = {1\over2}(w{\cal H}_\Phi\Phi + w\ov\Phi{\cal H}_{\ov\Phi}) .
\eeq
Acting with $D_\alpha$, using identity (\ref{4sc11}) and the field equations immediately leads to the improvement transformation (\ref{Imp3}) applied to the natural structure.

We may further improve the structure (\ref{Imp4}) to a Ferrara-Zumino supercurrent with $\chi_\alpha=0$.
This second improvement would lead to a supercurrent depending on the superfield $\Delta_{(w)}$,
\beq
\widehat J_{\alpha\dalpha} \quad\longrightarrow\quad
\widehat J_{\alpha\dalpha} + {1\over3}[D_\alpha,\ov D_\dalpha]\Delta_{(w)}.
\eeq
The content of the supercurrent structure (\ref{Imp4}) is however more intuitive, with the lagrangian 
superfield ${\cal H}$ defining the supercurrent superfield $\widehat J_{\alpha\dalpha}$ and the 
scale- and $R$-breaking superfields $\Delta_{(w)}$, $\widetilde\Delta_{(w)}$ and $\Xi_{(w)}$
defining the source superfields $\widehat X$ and $\widehat\chi_\alpha$.

\section{Anomalies and super-Yang-Mills theory}

Consider now super-Yang-Mills theory described by an effective Wilson lagrangian
${\cal L}_{W,\mu}$. This local lagrangian is obtained schematically by functional integration of the super-Yang-Mills 
lagrangian ${\cal L}_{SYM}$ with a low-energy cutoff $\mu$ kept in the perturbative regime.\footnote{The
lagrangian ${\cal L}_{W,\mu}$ is then used to calculate amplitudes with $\mu$ as UV cutoff.} 
With the linear superfield used as gauge-coupling field, we have {\it two} gauge-invariant superfields, $\hat L$ 
and $\widetilde\Tr{\cal WW}$. In principle, at two-derivative level, 
\beq
\label{SYM1}
{\cal L}_\mu = \Dint {\cal H} ( \hat L)
+ \Fint W(\widetilde\Tr{\cal WW}) + {\rm h.c.}
\eeq
Omitting fermionic terms\footnote{Supersymmetric contributions which vanish in a bosonic backgound.} generated
by higher-order terms in $\widetilde\Tr{\cal WW}$, the superpotential reduces to
\beq
\label{SYM2}
{A\over4} \Fint \widetilde\Tr{\cal WW} = {A\over2} \Dint \hat L
\eeq
up to a derivative $\partial_\mu(\ldots)$ and it can be absorbed in ${\cal H}$. At two derivatives then, ${\cal H}$ only depends 
on $\hat L$. 

We next identify the scalar $C = \hat L|_{\theta=0}$ with the gauge coupling defined at some arbitrary scale $M$:
\beq
\label{SYM3}
C= m^2 g^2(M),
\eeq
where the irrelevant mass scale $m$ keeps track of the mass dimension two of $C$.\footnote{Scale transformations 
act on $C$ and not on $m$. But $\mu$ and $M$ are actually energy or momentum scales on which scale
transformations act. } As a consequence, we use
\beq
\label{SYM4}
{\cal L}_{W,\mu} = m^2\Dint {\cal H}\left({\hat L\over m^2}\right) =
m^2 {\cal H}_C \, {\cal L}_{SYM} + \ldots
\eeq
and the natural supercurrent structure (\ref{CSF12}) reduces to 
\beq
\label{SYM5}
\begin{array}{rcl}
\ov D^\dalpha J_{\alpha\dalpha} &=& D_\alpha X + \chi_\alpha ,
\crbig
J_{\alpha\dalpha} &=& -4\,m^2 {\cal H}_L \widetilde\Tr ({\cal W}_\alpha e^{-{\cal A}} \ov {\cal W}_\dalpha e^{\cal A}) 
+ 2 \, m^2 {\cal H}_{LL}(\ov D_\dalpha\hat L)(D_\alpha\hat L)  ,
\crbig
X &=& 0,
\qquad\qquad\qquad\qquad
\chi_\alpha \,\,=\,\, m^2\ov{DD}D_\alpha({\cal H}-\hat L{\cal H}_L).
\end{array}
\eeq
It includes the Belinfante energy-momentum tensor and the $U(1)_{\widetilde R}$ current (\ref{CSF13}).

A perturbative expansion of the Wilson gauge coupling
\beq
\label{SYM6}
{1\over g^2_W(\mu)} = m^2 {\partial {\cal H} \over\partial C} = m^2 \, {\cal H}_C
\eeq
would indicate
\beq
\label{SYM7}
{\cal H} = \ln \hat L + \sum_{k\ge1} {c_k\over k} \left[{\hat L\over m^2}\right]^k
\eeq
where the numbers $c_k$ are the $k$-loop corrections which depend on $\mu/M$ and on the physical 
coupling $g^2(M)$. Notice that with this expansion, the quantum corrections to $\chi_\alpha$ appear at two loops.
This however holds under the assumption that the quantum correction in ${\cal H}$ admits a power expansion 
around $g^2=0$ and this is not what we will find.
We are interested in a derivation of ${\cal H}$ to all orders using two arguments. 
Firstly, it is known that the $\mu$--dependence of the Wilson coupling stops at one-loop \cite{SV, SV86}:
\beq
\label{SYM8}
\mu{d\over d\mu} {1\over g_W^2} = { b_0 \over 8\pi^2} \, ,
\eeq
where $b_0 = 3 C(G)$ is the coefficient of the one-loop $\beta$ function.\footnote{And $C(G)$ is the quadratic 
Casimir $C(G)\delta^{ab} = f^{acd}f^{bcd}$ in terms of structure constants.} This indicates that
\beq
\label{SYM9}
\begin{array}{rcl}
{\cal H} &=& \displaystyle
\ln\hat L + {3C(G)\over8\pi^2} \ln{\mu\over M} \, {\hat L\over m^2} + {\cal H}_{pert.}(\hat L/m^2) ,
\crbig \displaystyle
{1\over g_W^2(\mu)} &=& \displaystyle {1\over g_W^2(M)} + {3C(G)\over8\pi^2} \ln{\mu\over M},
\qquad\quad
{1\over g_W^2(M)} \,\,=\,\,  {m^2\over C} +  m^2{\partial\over\partial C} {\cal H}_{pert.}(\hat L/m^2),
\end{array}
\eeq
where the quantum correction ${\cal H}_{pert.}$ does not depend on $\mu$.
Secondly anomaly-mat\-ching in the Wilson effective lagrangian.
The outcome will be an algebraic derivation, from anomalies, of the all-order $\beta$ function for 
super-Yang-Mills theory originally obtained by Novikov, Shifman, Vainshtein and Zakharov (NSVZ) \cite{NSVZ}
from instanton calculations of the gaugino condensate and by Jones \cite{J} using Ward identity arguments.

Quantum anomalies affect the chiral $\widetilde R$--current and the trace of the energy-mo\-men\-tum tensor, or the 
dilatation current. The $U(1)_{\widetilde R}$ transformation of the gaugino $\lambda_\alpha$ with charge $q=3/2$ 
induces a one-loop chiral $\widetilde R$--gauge--gauge mixed anomaly:\,\footnote{Dots indicate terms 
needed by supersymmetry.}
\beq
\label{anom1}
\partial^\mu j_\mu^{(\lambda)} = {1\over16\pi^2} \, {3\over2}C(G)\, \widetilde\Tr \, F^{\mu\nu} \widetilde F_{\mu\nu}
+ \ldots ,
\qquad\qquad
j_\mu^{(\lambda)} = {3\over2} {1\over g_W^2(\mu)} \lambda\sigma^\mu\ov \lambda .
\eeq
Since $\partial^\mu j_\mu = - {3\over2}\Im f_X$ in the supercurrent structure, the anomaly adds a quantum
correction to the chiral source superfield
\beq
\label{anom2}
X_{(anomaly)} = - {1\over8\pi^2} \, C(G) \, \widetilde\Tr {\cal WW}.
\eeq
Comparing with $\widetilde X$ in (\ref{Imp4}), we can write
\beq
\label{anom3}
X_{(anomaly)} = -{4\over3}\,\widetilde\Delta_{(anomaly)}, \qquad\qquad
\widetilde\Delta_{(anomaly)} = {3\over32\pi^2} \, C(G)\,\widetilde\Tr {\cal WW}.
\eeq
And using the definition (\ref{Deltais}), the anomaly could be generated in an effective lagrangian with the 
$F$--term superpotential
\beq
\label{anom4}
W_{(anomaly)} =  {1\over32\pi^2} \, C(G)\,\widetilde\Tr {\cal WW} \Bigl[ \ln \widetilde\Tr {\cal WW}- 1 \Bigr]
\eeq
since $\widetilde\Tr {\cal WW}$ has $\widetilde R$--charge and scale dimension three.\footnote{Up to
an arbitrary term linear in $\widetilde\Tr{\cal WW}$ which would set its scale.} 
 At the perturbative level, as 
earlier indicated, the superpotential (\ref{anom4}) is fermionic. When gauginos condensate,
it gives rise to the Veneziano-Yankielowicz superpotential \cite{VY}.
Under a scale or $\widetilde R$ transformation with parameters $\beta$ and $\alpha$,
\beq
\label{anom5}
\delta \int d^4x\Fint W_{(anomaly)} + {\rm h.c.} =  {3C(G)\over32\pi^2} \, (\beta+ i\alpha) \,
\int d^4x \Fint \widetilde\Tr {\cal WW} + {\rm h.c.}
\eeq
This is precisely the variation induced by a formal rescaling $\mu\rightarrow e^{\beta+i\alpha}\mu$
of the Wilson scale in the lagrangian defined by expression (\ref{SYM9}). 
In this sense, the one-loop correction to ${\cal L}_{W,\mu}$ is a one-loop anomaly-matching term. 

Supersymmetry relates the contributions of $X_{(anomaly)}$ to $\partial^\mu j_\mu$ and to ${T^\mu}_\mu$:
\beq
\label{anom6}
\partial^\mu j_\mu = {3\,C(G)\over32\pi^2} \, \widetilde\Tr \, F^{\mu\nu} \widetilde F_{\mu\nu} +\ldots 
\qquad \longleftrightarrow \qquad
{T^\mu}_\mu = {3\,C(G)\over8\pi^2} \, {\cal L}_{SYM} + \ldots
\eeq
but the last equation is not the result predicted by the dilatation anomaly of a gaugino with scale dimension
$3/2$,\footnote{Gauginos are superpartners of gauge fields, their anomalous dimensions vanish.}
\beq
\label{anom7}
{T^\mu}_\mu = {1\over12\pi^2} \, {3\over2}C(G)\, {\cal L}_{SYM}  + \ldots
\eeq 
Hence, there is a residual anomaly
\beq
\label{anom7b}
{T^\mu}_\mu - {3C(G)\over8\pi^2} \, {\cal L}_{SYM} = - {C(G)\over4\pi^2}\,{\cal L}_{SYM} + \ldots
\eeq
This residual anomaly must be compensated in the Wilson effective lagrangian by renormalization-group 
invariance of the theory, in terms of the dependence on $M$ or $g^2(M)$ or $C$.

We can now use the real $\hat L$ and the source superfield $\chi_{(anomaly)\alpha}$ for this cancellation 
mechanism. We need
\beq
\label{anom8}
\chi_{(anomaly)\alpha} = - {C(G) \over8\pi^2} \, \ov{DD}D_\alpha \, \hat L.
\eeq
The last eq.~(\ref{SYM5}) indicates that this contribution is generated by the anomaly counterterm
\beq
\label{anom9}
{\cal H}_{(anomaly)} = {C(G)\over8\pi^2}\,{\hat L\over m^2} \Bigl[ \ln {\hat L\over m^2} - 1 \Bigr],
\eeq
up to an arbitrary invariant linear term.\footnote{Which is a one-loop term, see eq.~(\ref{SYM7}).}
This counterterm is the all-order correction ${\cal H}_{pert.}$
in the effective lagrangian which is then defined by
\beq
\label{anom10}
m^2{\cal H} = m^2\ln\hat L + {3C(G)\over8\pi^2} \ln{\mu\over M} \, \hat L 
+ {C(G)\over8\pi^2} \Bigl[\hat L\ln{\hat L\over m^2} - \hat L \Bigr] .
\eeq
Hence, the presence of the gauge coupling superfield $L$ takes care of the {\it different} values of the 
chiral $\widetilde R$ and ${T^\mu}_\mu$ anomalies, and produces the all-order correction ${\cal H}_{pert.}$.
In the lagrangian defined by eq.~(\ref{anom10}), the Wilson gauge coupling is
\beq
\label{anom11}
\begin{array}{rcl} \displaystyle
{1\over g^2_W(\mu)} &=& \displaystyle 
{m^2\over C} + {3C(G)\over8\pi^2} \ln{\mu\over M} + {C(G)\over8\pi^2} \ln{C\over m^2}
\crbig
&=& \displaystyle {1\over g^2(M)} + {3C(G)\over8\pi^2} \ln{\mu\over M} + {C(G)\over8\pi^2} \ln g^2(M),
\crbig \displaystyle
{1\over g^2_W(M)} &=& \displaystyle {1\over g^2(M)} + {C(G)\over8\pi^2} \ln g^2(M).
\end{array}
\eeq
Finally, since the reference energy scale $M$ is arbitrary, the renormalization-group equation 
\beq
\label{anom12}
M{d\over dM} g^2_W(\mu) = 0
\eeq
leads to the $\beta$ function
\beq
\label{anom13}
\beta (g^2) = M{d\over dM} g^2(M) =  {3C(G)\over8\pi^2}[m^4{\cal H}_{CC}]^{-1}
= - {g^4 \over 8\pi^2} \, {3\,C(G) \over 1 - {g^2\over8\pi^2} C(G) } 
\eeq
with $g^2(M)=C/m^2$. Hence, the NSVZ $\beta$ function (\ref{anom13}) \cite{NSVZ}, with its two coefficients, follows
from the matching of the $U(1)_{\widetilde R}$ and dilatation anomalies and, in the Wilson effective lagrangian,
from the one-loop running which defines the Wilson coupling $g_W^2$. The important point is the existence
of two gauge-invariant superfields, the chiral $\widetilde\Tr{\cal WW}$, as usual, and the real $L-2\Omega$
which appears when the linear superfield is used as gauge coupling field. These two superfields are in natural 
relation with the two anomaly superfields $X$ and $\chi_\alpha$ of the supercurrent structure. Notice also
that holomorphicity is entirely absent in this discussion of ${\cal N}=1$ super-Yang-Mills, in contrast
with ${\cal N}=2$ with its K\"ahler scalar manifold.
A similar line of reasoning can be followed to derive the effective action for gaugino condensation, using
both counterterms (\ref{anom4}) and (\ref{anom9}).\footnote{See refs.~\cite{BDQQ, AADH}.}

In simple string compactifications to four dimensions with ${\cal N}=1$ supersymmetry, the linear
multiplet $L$ describes the dilaton and is then, naturally, the gauge coupling and loop-counting superfield
\cite{CFV}. Its role in anomaly cancellation of K\"ahler anomalies \cite{LCO} and in particular in the derivation 
of heterotic gauge threshold corrections \cite{DFKZ} has been established long ago. As a sequel, in the 
framework of conformal supergravity, a (somewhat obscure) derivation from anomalies of the NSVZ $\beta$ 
function with the linear gauge coupling field has already been given in ref.~\cite{DFKZ2}. In this approach, the 
renormalization-group behaviour is the response of the theory to a rescaling of the compensating field for dilatation 
symmetry \cite{GW, GGRS}. This section proposes a derivation in the simpler framework of global ${\cal N}=1$ 
supersymmetry, based on similar arguments and using supercurrent structures.

With constant gauge coupling, the all-order results (\ref{anom11}) have been obtained by
Shifman and Vainshtein \cite{SV}. The importance and a calculation of the residual anomaly (\ref{anom7b})
for the NSVZ $\beta$ function has been given with much clarity by Arkani-Hamed and Murayama \cite{AHM}.

Strictly speaking, theory (\ref{SYM4}) does not have an axion: the helicity zero superpartner of $C$ is
the antisymmetric tensor $B_{\mu\nu}$ with gauge invariance $\delta B_{\mu\nu} = \partial_\mu\Lambda_\nu
- \partial_\nu\Lambda_\mu$ and $L$ only includes the gauge-invariant curl 
\beq
\label{bmunu1}
H_{\mu\nu\rho} = 3\,\partial_{[\mu}B_{\nu\rho]} - \omega_{\mu\nu\rho}.
\eeq
The antisymmetric tensor couples to the gauge Chern-Simons form $ \omega_{\mu\nu\rho}$ and
the effective lagrangian does not have a perturbative dependence on a vacuum angle even
if the gauge coupling has an all-order expansion. The antisymmetric tensor is dual to a pseudoscalar $\sigma$
with axionic shift symmetry and universal coupling 
$$
{1\over4} \, \sigma \, \widetilde\Tr F_{\mu\nu}\widetilde F^{\mu\nu}
$$
for all functions ${\cal H}$: the quantum corrections to ${\cal H}$ appear in the kinetic lagrangian
$- ({\cal H}_{CC})^{-1} \, (\partial_\mu\sigma)(\partial^\mu\sigma)$. It is admissible to work with 
$C$ and $\sigma$ but the resulting chiral supermultiplet is not in a K\"ahler basis and
supersymmetry variations explicitly depend on the function ${\cal H}$ defining the lagrangian. With
${\cal H}$ as given in expression (\ref{anom10}), the Legendre transformation 
turning $C$ into the standard superpartner of $\sigma$ in a k\"ahlerian chiral multiplet cannot be 
analytically solved and information would be lost in an approximate treatment.

\section*{Acknowledgements}

I wish to thank the organizers of a brilliant Planck conference at Ioannina University, Jelle Hartong and, 
at earlier stages of this work, Nicola Ambrosetti and Daniel Arnold for collaboration. Parts of my
work have been performed during several stays at the Ecole Normale Sup\'erieure, Paris, the Ecole Polytechnique,
Palaiseau and the NCTS at National Tsing Hua University, Hsinchu, Taiwan. Hospitality and discussions 
with many colleagues are gratefully acknowledged.


\end{document}